\documentclass{article}

\usepackage{PRIMEarxiv}

\usepackage[utf8]{inputenc} 
\usepackage[T1]{fontenc}    
\usepackage{hyperref}       
\usepackage{url}            
\usepackage{booktabs}       
\usepackage{amsfonts}       
\usepackage{nicefrac}       
\usepackage{microtype}      
\usepackage{lipsum}
\usepackage{fancyhdr}       
\usepackage{graphicx}       
\graphicspath{{media/}}     

\usepackage{booktabs}
\usepackage{tabularx}
\usepackage{subcaption}
\usepackage{enumitem}
\usepackage[most]{tcolorbox}
\usepackage[ruled,vlined,linesnumbered]{algorithm2e}
\usepackage{multirow}
\usepackage[table]{xcolor}
\definecolor{bestbg}{HTML}{DCE7DC}

\definecolor{answerbg}{HTML}{F4F7FB}
\definecolor{answerline}{HTML}{5B6E8C}
\newtcolorbox{answerbox}{
  enhanced, colback=answerbg, colframe=answerline, boxrule=0.6pt, arc=4pt,
  left=10pt, right=10pt, top=7pt, bottom=7pt, boxsep=0pt,
  before skip=8pt, after skip=8pt
}

\newcommand{\toolname}[1]{\texttt{[#1]}}
\newcommand{\gsname}[1]{\texttt{#1}}
\newcommand{\bugname}[1]{\textsf{#1}}

\title{REINS: Agent-guided Thread Scheduling for Concurrency Bug Detection and Reproduction}

\author{
  Yu Zhang \\
  The University of Queensland \\
  \texttt{yu.zhang3@uq.edu.au} \\
   \And
   Corina S. Pasareanu \\
  Carnegie Mellon University \\
  \texttt{pcorina@andrew.cmu.edu} \\
   \And
  Guowei Yang \\
  The University of Queensland\\
  \texttt{guowei.yang@uq.edu.au} \\
}

\begin{document}
\maketitle

\begin{abstract}
Concurrency bugs manifest only under specific thread interleavings, making them hard to detect and reproduce.
Existing controlled concurrency testing techniques take over thread scheduling to explore the interleaving space.
As the space grows rapidly with program scale, finding interleavings that can trigger bugs within a limited budget remains challenging.
The program's concurrency semantics can reveal which variable accesses and synchronization operations under specific interleavings lead to unexpected program behavior, and thus help focus the exploration on bug-relevant interleavings. 
However, prior techniques have not fully exploited such semantic information.
In this paper, we present \textsc{Reins}, an agentic approach for detecting and reproducing concurrency bugs.
At its core, \textsc{Reins} guides thread scheduling with the agent's reasoning about the concurrency semantics and known concurrency bug patterns.
\textsc{Reins} starts with runtime observations and combines them with static information to reason about the program's concurrency behavior and formulate hypotheses about potential bugs.
A hypothesis translates into a set of constraints on event orders and variable states, guiding a controlled scheduler toward interleavings likely to trigger the bug.
\textsc{Reins} draws conclusions from the execution results and keeps refining the constraints until the bug is triggered or the hypothesis is refuted.
Compared with existing techniques on three public benchmarks, \textsc{Reins} detects more bugs within the same time budget.
Furthermore, \textsc{Reins} found six previously unknown concurrency bugs in real-world projects, three of which have been confirmed by the developers.
\end{abstract}


\section{Introduction}
Concurrency bugs are notoriously difficult to detect and reproduce~\cite{musuvathi2008finding,thomson2016concurrency}. Due to the non-determinism of thread scheduling, they may manifest only under specific thread interleavings, making them difficult to expose during testing.
Once triggered in production, they can lead to severe consequences, such as program crashes, data corruption, and security vulnerabilities~\cite{lu2008learning}.
Traditional stress testing, which relies on massive repeated executions and random perturbations, struggles to expose concurrency bugs in the vast thread-interleaving space~\cite{edelstein2002multithreaded,musuvathi2008finding}.
Controlled concurrency testing (CCT) takes over thread scheduling to make the exploration and reproduction of interleavings controllable, and has therefore become one of the prevailing approaches to concurrency bug detection and reproduction. 
CCT mainly proceeds through systematic enumeration~\cite{musuvathi2007iterative, musuvathi2008finding,emmi2011delay} or randomized sampling with probabilistic guarantees~\cite{burckhardt2010randomized,yuan2018partial}. 
Recent work further improves the scheduling strategies, such as periodical scheduling~\cite{wen2022controlled} and selectively uniform sampling~\cite{zhao2025selectively}, or introduces ideas from reinforcement learning~\cite{mukherjee2020learning} and greybox fuzzing~\cite{wolff2024greybox} to guide scheduling decisions.
Despite these advances, bugs with narrow triggering windows and specific state preconditions remain difficult to expose within a limited budget. 
A key limitation of existing approaches is their insufficient exploitation of concurrency semantics of the program, 
e.g., which combinations of accesses are dangerous under the program logic, and in what state the program must be for a bug to be triggered. 
Such information is hard to obtain at the schedule level where these approaches operate.
Recently, LLM-based agents have been applied to a wide range of software testing tasks~\cite{wang2024software,zhu2026locus,pu2026patch,wang2024hits,cai2026concovup}, demonstrating strong capabilities in program comprehension and reasoning.
These advances motivate us to investigate whether an agent's reasoning about concurrency semantics can guide thread scheduling to more effectively trigger concurrency bugs.

However, effectively applying LLM agents to concurrency testing presents several key challenges, which we summarize as follows.
\textbf{C1: Incomplete understanding of concurrency semantics.} LLM agents struggle to comprehend the concurrency semantics~\cite{huang2026concur,chen2023data}, especially in large programs where a concurrency bug often spans multiple threads, functions, and files. Constrained by their context, the agents may not be able to piece together a reliable global view and locate bug-relevant operations in massive code bases. 
\textbf{C2: The gap between concurrency semantics and thread scheduling.} 
The agent reasons at the level of program semantics, focusing on the roles of variables and operations and on which thread interactions may lead to unexpected behavior. On the other hand, the scheduler must decide which thread to run next at every scheduling point during execution, and a single run may involve thousands of such points. 
Requiring the agent to make every scheduling decision is impractical, while its high-level semantic reasoning cannot be directly translated into concrete thread choices at individual scheduling points.
\textbf{C3: Unreliable agent output.} The agent may produce plausible yet incorrect reasoning~\cite{turpin2023language,huang2025survey}, and may even hallucinate variables and code locations that do not exist, so its claims do not directly constitute evidence.

In this paper, we propose \textsc{Reins}, an agentic approach that addresses the above challenges for detecting and reproducing concurrency bugs in shared-memory multithreaded programs. 
The key insight of \textsc{Reins} is to turn the agent's reasoning about the concurrency semantics into constraints that the scheduler recognizes and enforces, thereby guiding thread scheduling toward the interleavings that are likely to trigger bugs, while the results of the controlled executions in turn serve as the evidence to refine the agent's hypotheses.
Specifically, \textsc{Reins} introduces the following features to address the three challenges.
For \textbf{C1}, \textsc{Reins} starts from the shared-variable accesses observed in real executions. They reflect the actual interactions between threads and, although covering limited execution behavior, they help the agent to initially locate the key variables and operations. 
Starting from runtime information, the agent selectively examines relevant source code and queries the corresponding calling contexts and path conditions through static analysis to supplement information not captured by runtime observations. This enables the agent to reason about concurrency semantics without exhaustively exploring a large codebase. Based on this reasoning, the agent formulates \emph{bug hypotheses}, each characterizing the program interactions that may lead to a specific concurrency bug pattern and its resulting failure.
For \textbf{C2}, \textsc{Reins} bridges the agent's reasoning and the scheduler's execution with the \emph{schedule plan}. A schedule plan is a set of constraints on event orders and variable states, involving only the few key events and variables required to trigger a bug, which is the granularity the agent can express precisely at the semantic level. The agent derives the constraints from a hypothesis and writes them into a plan for the scheduler to execute. The scheduler controls the execution according to the plan at the scheduling points relevant to the constraints, and decides the rest at random.
For \textbf{C3}, \textsc{Reins} checks the agent's output and takes execution results as evidence. The events and variables in a plan proposed by the agent are validated against the program before the plan is executed by the controlled scheduler. The conclusions of \textsc{Reins} rest on real executions. For an execution that triggers the bug, its scheduling trace is recorded so that the bug can be deterministically replayed. For the executions that do not trigger the bug, the results distinguish whether the interleaving did not happen as planned or happened without the expected failure, and the agent then revises the plan or the hypothesis accordingly and iterates with subsequent executions.

\textsc{Reins} proceeds iteratively through three phases: \emph{comprehend}, \emph{attempt}, and \emph{assess}.
The comprehend phase builds the understanding of the concurrency semantics of the program and proposes a set of bug hypotheses as \emph{candidates}. 
The attempt phase derives schedule plans for the candidates and executes them to trigger the hypothesized bugs.
The assess phase analyzes the results, propagates conclusions to related candidates, and decides whether the next iteration returns to the comprehend phase for further collection or moves on to the attempt phase for the remaining candidates. \textsc{Reins} continues until all candidates are settled or the time budget is exhausted.

We evaluate \textsc{Reins} on three public benchmarks~\cite{thomson2014concurrency, cai2019detecting, liang2023racebench}, and compare it with representative controlled scheduling techniques~\cite{wolff2024greybox,zhao2025selectively,burckhardt2010randomized,thomson2016concurrency}. \textsc{Reins} detected more bugs than all baselines across the three benchmarks. In particular, on the largest benchmark, RaceBench~\cite{liang2023racebench}, \textsc{Reins} detected 38 more bugs than the best-performing baseline~\cite{zhao2025selectively} within the same time budget. 
In addition, on two real-world projects, \textsc{Reins} detected six previously unknown concurrency bugs, three of which have been confirmed by their developers. 

In summary, this paper makes the following contributions:
\begin{itemize}
\item We propose agent-guided thread scheduling, an approach that leverages the agent's reasoning about the concurrency semantics to guide controlled scheduling toward bug-relevant schedules.
\item We design and implement \textsc{Reins}, an agentic system for detecting and reproducing concurrency bugs, which integrates agent reasoning with controlled scheduling and iterates through the comprehend, attempt, and assess phases.
\item We evaluate \textsc{Reins} against existing techniques on public benchmarks, where it detects more bugs within the same time budget, demonstrating its effectiveness. \textsc{Reins} further identifies six previously unknown bugs in real-world projects.
\end{itemize}

\section{Overview}
\label{sec:overview}
In this section, we first use a motivating example to illustrate why concurrency bugs are difficult to trigger, and then provide an overview of how \textsc{Reins} guides thread scheduling to detect such bugs.

\begin{figure}[t]
  \centering
  \includegraphics[width=\textwidth]{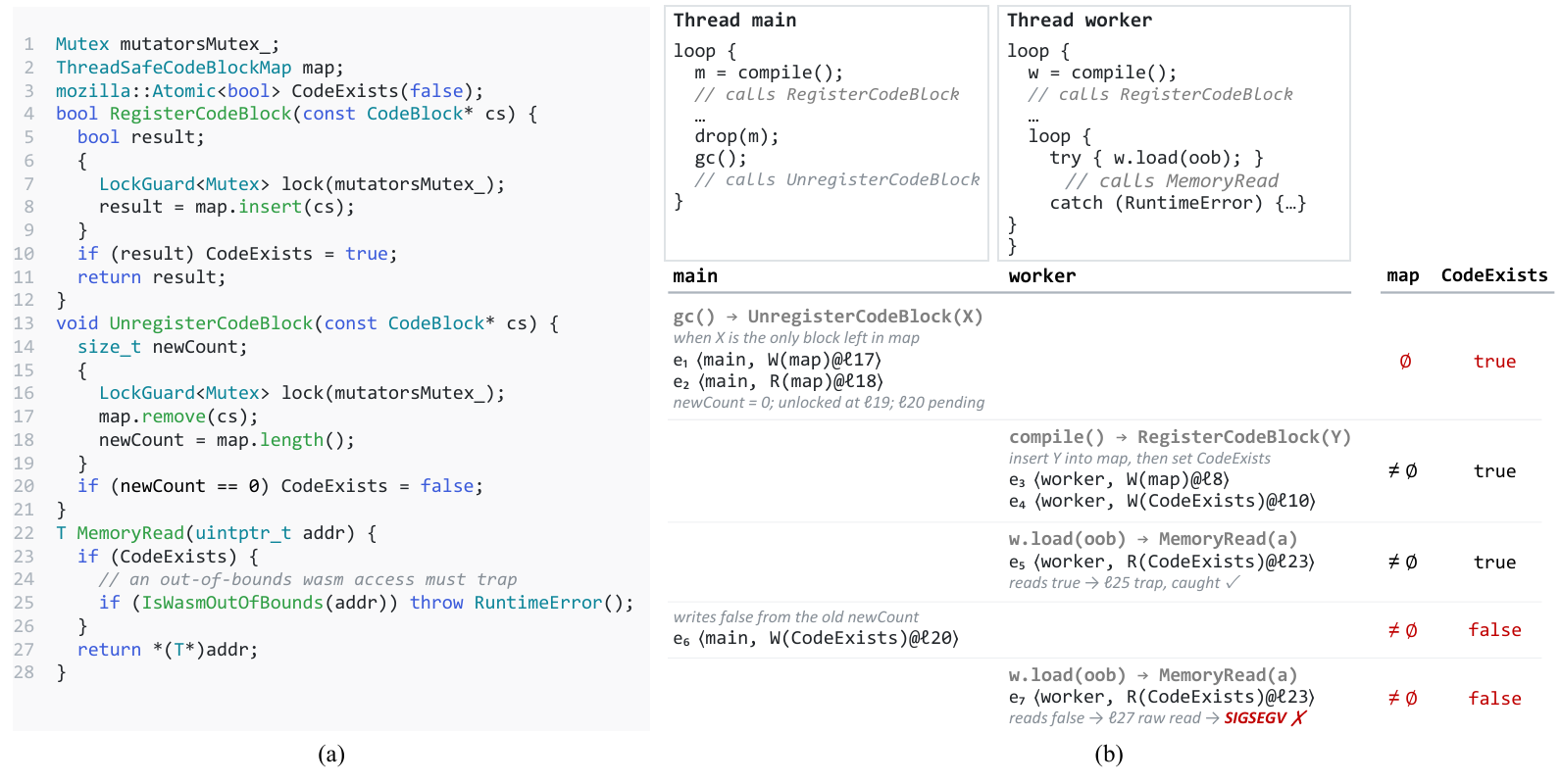}
 \caption{An atomicity violation from Mozilla SpiderMonkey that ends in a crash on an out-of-bounds access. \texttt{CodeExists} caches whether \texttt{map} is non-empty, but the two are not updated atomically. (a) The simplified key code snippet. (b) The thread interleaving that triggers the bug, with red marking the states where \texttt{map} and \texttt{CodeExists} disagree.}
  \label{fig:motivation}
\end{figure}

\textbf{Motivating Example.}
Figure~\ref{fig:motivation} shows an atomicity violation from Mozilla SpiderMonkey~\cite{spidermonkey}, the JavaScript and WebAssembly (wasm) engine of Firefox.
An out-of-bounds wasm memory access must be turned into a \texttt{RuntimeError} that JavaScript can catch, rather than crashing the process.
SpiderMonkey records the current wasm code blocks in a lock-protected \texttt{map}, caches whether \texttt{map} is non-empty in the atomic flag \texttt{CodeExists}, and performs the bounds check in \texttt{MemoryRead} only when \texttt{CodeExists} is true (lines 23--25).
Critically, a false \texttt{CodeExists} while \texttt{map} is non-empty causes \texttt{MemoryRead} to skip the required bounds check.
However, \texttt{UnregisterCodeBlock} computes the remaining count while holding the lock but updates \texttt{CodeExists} only after releasing it (lines 16--20).
As shown in Figure~\ref{fig:motivation}(b), the main thread removes the last code block and observes \texttt{newCount = 0} ($e_1$, $e_2$), but is delayed before updating the flag.
Meanwhile, the worker thread registers a new code block and sets \texttt{CodeExists} to \texttt{true} ($e_3$, $e_4$), restoring a consistent state in which \texttt{map $\neq \emptyset$} and \texttt{CodeExists = true}.
Its first out-of-bounds access is therefore correctly trapped ($e_5$).
The main thread then resumes and writes \texttt{CodeExists = false} based on its stale \texttt{newCount} ($e_6$), overwriting the worker's update and leaving \texttt{map $\neq \emptyset$} while \texttt{CodeExists = false}.
A subsequent out-of-bounds access by the worker reads this false flag, skips the bounds check, and directly accesses the invalid address ($e_7$), resulting in \texttt{SIGSEGV}.
Triggering this bug requires several ordering constraints to hold at once. The main thread must be delayed between observing \texttt{newCount = 0} and writing the flag, the worker must complete a registration within this interval, and its subsequent out-of-bounds access must occur before any later registration corrects the flag. Such opportunities are rare. The window opens only when the last code block is unregistered, and it lasts only a single statement, during which the worker must happen to complete a registration. Such an interleaving is therefore unlikely to arise under natural scheduling and calls for explicit control over thread execution.

\textbf{Controlled Scheduling.}
Controlled concurrency testing~\cite{musuvathi2007iterative, musuvathi2008finding,emmi2011delay,burckhardt2010randomized,yuan2018partial} takes over thread scheduling by inserting scheduling points at key operations, such as shared-memory accesses and synchronization primitives, where the scheduler selects a runnable thread to execute. An execution is thus represented by a sequence of scheduling decisions, which can be recorded to deterministically reproduce a bug-triggering execution.
Such techniques can cover the buggy interleaving in Figure~\ref{fig:motivation} in theory. However, real-world programs contain numerous scheduling points, and the scheduler does not know which decisions are relevant to the bug, so reaching the target interleaving remains challenging for bugs with narrow triggering windows and multiple ordering constraints.

\begin{figure*}[t]
  \centering
  \includegraphics[width=\textwidth]{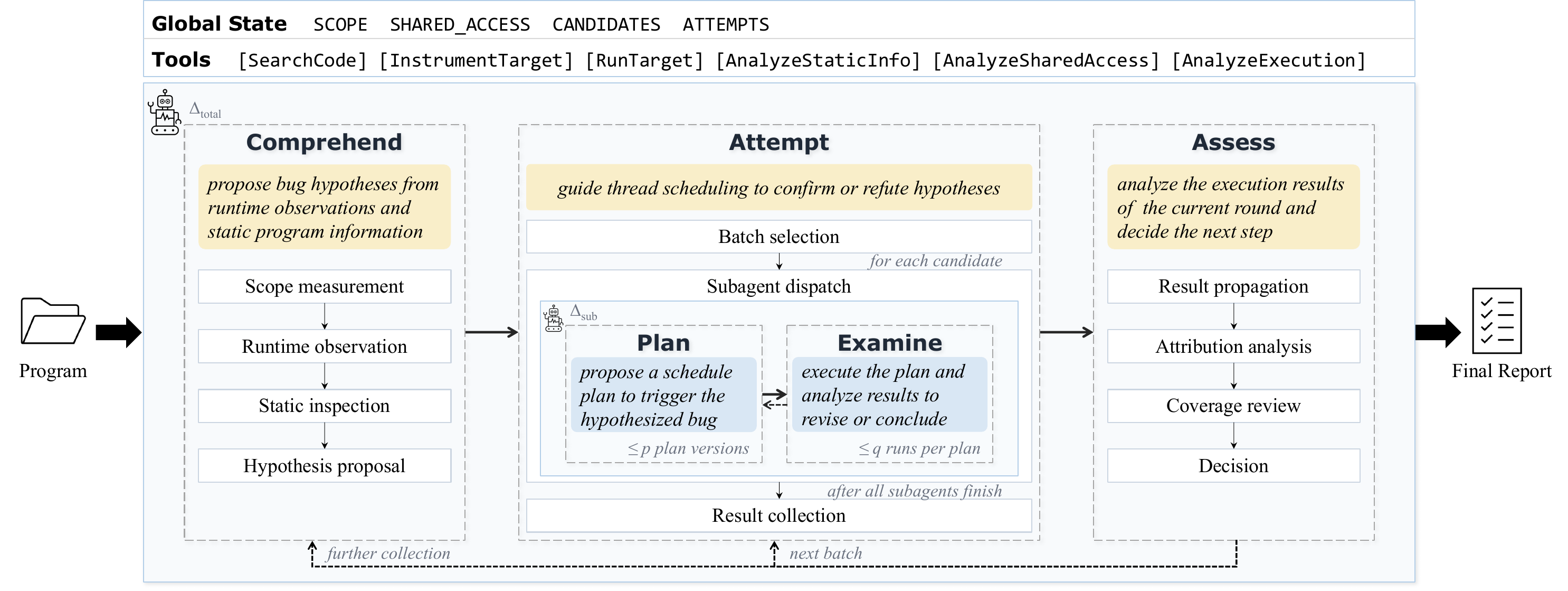}
\caption{Overview of \textsc{Reins}. The agent runs in three phases within a total time limit $\Delta_{\mathrm{total}}$, and each subagent runs within a time limit $\Delta_{\mathrm{sub}}$, with at most $p$ plan versions and $q$ runs per plan.}
  \label{fig:framework}
\end{figure*}

\textbf{Agent-guided Thread Scheduling.}
Triggering the bug in Figure~\ref{fig:motivation} depends on only a few scheduling decisions, yet nothing at the schedule level tells them apart from the numerous others. \textsc{Reins} combines runtime observations and static information to help the agent propose a bug hypothesis: \texttt{CodeExists} is updated outside the lock and may overwrite the \texttt{true} just set by another thread with \texttt{false}, which in turn lets an out-of-bounds access skip the check and crash. From the hypothesis, \textsc{Reins} further derives the event orders and variable states required to trigger the bug. In terms of order, the worker must complete a registration ($e_4$) and enter its memory-access loop ($e_5$) before the main thread writes the stale \texttt{false} ($e_6$), and the worker must then perform another out-of-bounds access ($e_7$), so that $e_7$ reads the \texttt{false} written by $e_6$ rather than a \texttt{true} corrected by a later registration. In terms of variable state, \texttt{CodeExists} must be \texttt{true} when $e_6$ occurs, so that $e_6$ overwrites the valid value just set by $e_4$, the inconsistent state with \texttt{map} non-empty and \texttt{CodeExists} false arises, and $e_7$ skips the check and crashes.
These requirements are expressed as constraints for the scheduler, which prioritizes or delays threads at the relevant scheduling points to satisfy them and decides the remaining points at random. 
The execution results feed back whether the intended interleaving is reached and whether the bug is triggered, and the agent refines the constraints or revises the hypothesis accordingly.
Figure~\ref{fig:framework} shows the workflow of \textsc{Reins}, driven by a lead LLM agent\footnote{Unless otherwise stated, ``the agent'' in the rest of the paper refers to the lead agent.} that iterates in \textit{rounds} through the comprehend, attempt, and assess phases. The first round starts from the comprehend phase, which proposes bug hypotheses as candidates. In the attempt phase, the agent dispatches a subagent for each selected candidate to propose and execute plans. Every round ends with the assess phase, which consolidates the results of the round and decides which phase the next round starts from. We formalize the related notions in Section~\ref{sec:preliminary} and detail \textsc{Reins} in Section~\ref{sec:method}.

\section{Preliminary}
\label{sec:preliminary}
In this section, we introduce the foundational notions and relevant definitions.

\textbf{Basic Notions.}
We use the symbol $P$ to denote the target concurrent program, the symbol $c$ to denote a source file in $P$, and $\ell$ to denote the line number in source code.
The symbols $F$ ($f \in F$) and $V$ ($v \in V$) denote the set of functions and the set of variables, respectively, and $T$ ($t \in T$) denotes the set of threads spawned during execution.
A \emph{variable access site} $\tau = \mathit{op}(v)@c{:}\ell$ denotes a read or write operation $\mathit{op} \in \{\mathsf{R}, \mathsf{W}\}$ on variable $v$ at line $\ell$ of source file $c$.
An \emph{event} $e = \langle t, \tau \rangle$ represents thread $t$ performing the access $\tau$ during execution.

\textbf{Definition 1.}
The \emph{observed access summary} of a variable $v$ is $\sigma_v = \langle v, [\langle e_1, L_1 \rangle, \langle e_2, L_2 \rangle, \ldots] \rangle$, which records the events on $v$ in the order observed over a set of executions, where $L_i$ is the set of locks held by thread $t$ when the event $e_i = \langle t, \tau \rangle$ occurs. 

\textbf{Definition 2.}
A \emph{bug hypothesis} is a conjecture proposed by the agent about a potential concurrency bug, denoted as $h = \langle V_h, \mathcal{T}_h, \delta \rangle$, where $V_h$ is the set of variables involved, $\mathcal{T}_h$ is the set of access sites on which the hypothesis rests, and $\delta$ is a natural-language description of the hypothesis, i.e., which accesses interleave in which way to go wrong, and its expected failure.

\textbf{Definition 3.}
A \emph{schedule plan} consists of a set of constraints, denoted as $\pi = \langle S, \Phi \rangle$, where $S = [e_1 \prec e_2 \prec \cdots \prec e_n]$ is a sequence of \emph{ordering constraints}, each $e \prec e'$ requiring that event $e$ occur before event $e'$; and $\Phi = \{(e, \phi)\}$ is a set of \emph{state constraints}, each pair $(e, \phi)$ requiring that $\phi$ hold when event $e$ occurs, where $\phi = \bigwedge_i \varphi(v_i)$ is a conjunction of predicates, each $\varphi(v_i)$ restricting the value of an involved variable $v_i$.

\textbf{Example 1.}
For the bug in Figure~\ref{fig:motivation}(a), the schedule plan that eventually triggers the bug is $S = [e_4 \prec e_5 \prec e_6 \prec e_7]$ and $\Phi = \{(e_6, \mathtt{CodeExists} = \mathsf{true})\}$, with events as labeled in Figure~\ref{fig:motivation}(b). The ordering constraints require the four events to occur in sequence, and the state constraint requires \texttt{CodeExists} to be \texttt{true} when $e_6$ occurs.

\textbf{Definition 4.} 
The \emph{execution result} of a single run of a schedule plan $\pi$ is a tuple $x = \langle \epsilon, \kappa_S, \kappa_\Phi, \theta, \gamma \rangle$. The termination result of the run is denoted by $\epsilon \in \{\textsf{normal}, \textsf{triggered}, \textsf{timeout}, \textsf{other}\}$, indicating that the run exits normally, triggers the expected bug, times out without triggering it, or fails for a reason other than the expected bug. The symbol $\kappa_S \in \{0, \ldots, n\}$ is the length of the prefix of $S = [e_1 \prec \cdots \prec e_n]$ satisfied in order, i.e., $e_1, \ldots, e_{\kappa_S}$ occur in the planned order while $e_{\kappa_S+1}$ does not, with $\kappa_S = n$ meaning all are satisfied. Similarly, $\kappa_\Phi \subseteq \Phi$ is the set of state constraints $(e, \phi)$ whose $\phi$ never holds when $e$ is ready to occur, being $\emptyset$ if all state constraints are satisfied. If $\kappa_S = n$ and $\kappa_\Phi = \emptyset$, the run is \emph{engaged} as planned. The scheduling trace of the run is denoted by $\theta: \{1, 2, \ldots\} \to T$, mapping the $i$-th scheduling point to the thread $\theta(i) \in T$ chosen to execute there, which allows the run to be deterministically replayed. The \emph{new coverage} of the run is denoted by $\gamma = \langle C_\gamma, \mathcal{T}_\gamma \rangle$, where $C_\gamma$ is the set of source files and $\mathcal{T}_\gamma$ the set of access sites reached in this run but not observed before.
Executing $\pi$ multiple times yields the result sequence $X_\pi = [x_1, x_2, \ldots]$, whose aggregation $\hat{X}_\pi$ summarizes the number of runs and engaged runs, the count of each $\epsilon$, the distributions of $\kappa_S$ and $\kappa_\Phi$, the scheduling trace $\theta$ of a triggering run, and the merged $\gamma$.

\textbf{Problem Statement.} Given a shared-memory multithreaded program $P$ with its test entries and inputs, we aim to detect and reproduce concurrency bugs in $P$ within a time budget $\Delta_{\mathrm{total}}$ through agent-guided thread scheduling. This work targets the common concurrency bug patterns~\cite{lu2008learning,savage1997eraser}, including data races, atomicity violations, order violations, and deadlocks, that lead to observable execution failures, such as crashes, assertion failures, and hangs. Silent concurrency errors that produce wrong results while the program terminates normally are not considered currently.

\section{Methodology}
\label{sec:method}
In this section, we introduce \textsc{Reins} in detail.
As shown in Figure~\ref{fig:framework}, \textsc{Reins} takes the target program as input and proceeds in rounds through three phases within the time budget $\Delta_{\mathrm{total}}$, maintaining a set of global states and equipped with several tools.
Each round verifies a batch of candidates, dispatching one subagent per candidate. A subagent works within a time limit $\Delta_{\mathrm{sub}}$, proposing at most $p$ plan versions and running each version at most $q$ times. 
At the end, \textsc{Reins} generates a final report of the detected bugs, each with its scheduling trace $\theta$ for reproduction.
We first introduce the global states and the execution infrastructure, and then describe the three phases in turn.

\subsection{Global States and Execution Infrastructure}
\label{sec:gs_tools}
Throughout the process, \textsc{Reins} maintains four global states and provides the agent with an execution infrastructure of tools and a controlled scheduler. We describe them as follows.

\textbf{Global States.} The four global states are shown at the top of Figure~\ref{fig:framework}.
\gsname{SCOPE} records the set of source files $\{c\}$ reached during execution, measured initially in the comprehend phase and extended in the attempt phase with newly reached files.
\gsname{SHARED\_ACCESS} records the observed access summaries $\{\sigma_v\}$ of all shared variables, initially collected in the comprehend phase and extended in the attempt phase as new access sites or events are observed.
\gsname{CANDIDATES} records all bug hypotheses proposed by the agent as candidates, i.e., a set of triples $\langle h, \mathit{status}, \mathit{settledBy} \rangle$. Each candidate is tagged with a \textit{status}: \textsf{pending} for candidates not yet verified, \textsf{succeeded} for those triggered, \textsf{refuted} for those shown by evidence not to be bugs, and \textsf{uncertain} for those tried but neither triggered nor refuted. The field \textit{settledBy} records whether the status is settled by the candidate's own attempts (\textsf{own}) or by a related candidate (\textsf{related}), and is initialized to $\bot$.
\gsname{ATTEMPTS} records the attempts on each candidate as a map $h \mapsto [\langle \mathit{round}, A \rangle, \ldots]$, where each attempt is made in one round and $A = [\langle \pi_1, \hat{X}_{\pi_1} \rangle, \langle \pi_2, \hat{X}_{\pi_2} \rangle, \ldots]$ is the sequence of schedule plans tried in that round with their aggregated execution results.

\begin{table}[t]
  \caption{Tools in \textsc{Reins}.}
  \label{tab:tools}
  \small
  \begin{tabularx}{\columnwidth}{@{}l X@{}}
    \toprule
    \textbf{Tool} & \textbf{Description} \\
    \midrule
    \toolname{SearchCode} & Search source code by name, pattern, or location \\
    \toolname{InstrumentTarget} & Instrument and build binaries for recording or scheduling: \texttt{recorder} $\mid$ \texttt{scheduler} \\
    \toolname{RunTarget} & Run the target in one of the modes: \texttt{natural} $\mid$ \texttt{collect} $\mid$ \texttt{control} $\mid$ \texttt{replay} \\
    \toolname{AnalyzeStaticInfo} & Static analysis for the requested information: \texttt{callGraph} $\mid$ \texttt{pathCond} $\mid$ \texttt{sharedVar} \\
    \toolname{AnalyzeSharedAccess} & Build \gsname{SHARED\_ACCESS} and analyze access profiles and conflicting pairs \\
    \toolname{AnalyzeExecution} & Aggregate the execution results of a plan \\
    \bottomrule
  \end{tabularx}
\end{table}

\textbf{Tools.} Table~\ref{tab:tools} lists the six tools in \textsc{Reins}.
\toolname{SearchCode} retrieves source code by function name, regular expression, or file and line range.
\toolname{InstrumentTarget} instruments the variable access sites in the target with a unified probe macro and compiles it into two binaries, covering all sites in \gsname{SCOPE} by default or a specified subset on request. In the \texttt{recorder} binary, the probe records the executing thread, the access site, and the lock set held at that moment. In the \texttt{scheduler} binary, the probe yields control to the scheduler so that each site becomes a scheduling point.
\toolname{RunTarget} runs the target in four modes. The \texttt{natural} mode runs the program as is. The \texttt{collect} mode runs the recorder binary and collects the access traces. The \texttt{control} mode first validates whether the variables and locations in a given plan $\pi$ actually exist. A valid plan is executed by running the scheduler binary under the controlled scheduler, and an invalid one is returned with the reason. The \texttt{replay} mode deterministically replays a run under the controlled scheduler according to a given $\theta$.
\toolname{AnalyzeStaticInfo} performs static analysis on the Clang AST. The \texttt{callGraph} mode returns the callers and callees of a function and the call chains reaching a site. The \texttt{pathCond} mode extracts the branch conditions dominating an access site, and the \texttt{sharedVar} mode identifies the shared variables and their access sites.
\toolname{AnalyzeSharedAccess} analyzes the accesses collected at runtime. It first identifies the shared variables from the traces, i.e., the variables accessed by multiple threads, and aggregates their accesses into \gsname{SHARED\_ACCESS}. It then reports how many threads and sites access each variable, and identifies the \emph{conflicting access pairs} on a given variable, where two accesses form a pair if they come from different threads with at least one write, hold no common lock, and are not ordered by thread creation or join, together with the number of executions in which each condition holds and the distribution of their observed orders.
\toolname{AnalyzeExecution} aggregates the execution results $X_\pi$ of all runs of a plan into $\hat{X}_\pi$.

\begin{algorithm}[t]
\caption{Scheduling decision of the controlled scheduler in the \texttt{control} mode.}
\label{alg:sched}
\KwIn{plan $\pi = \langle S, \Phi \rangle$; runnable threads $T_{\mathit{rdy}}$; the length $s$ of the satisfied prefix of $S$}
\KwOut{the next thread $t^*$ to run}
partition $T_{\mathit{rdy}}$ into $T_{\mathit{next}}$, $T_{\mathit{held}}$, and $T_{\mathit{free}}$\;
\eIf{$T_{\mathit{next}} \neq \emptyset$}{
  $t^* \leftarrow$ a thread in $T_{\mathit{next}}$; \quad $s \leftarrow s+1$\;
}{
  \eIf{$T_{\mathit{free}} \neq \emptyset$}{
    $t^* \leftarrow$ a thread drawn from $T_{\mathit{free}}$ at random\;
  }{
    $t^* \leftarrow$ a thread forcibly released from $T_{\mathit{held}}$, not advancing $s$\;
  }
}
\If{\textnormal{$s$ has not advanced within a timeout}}{abandon the remaining constraints\;}
append $t^*$ to $\theta$; \Return $t^*$\;
\end{algorithm}

\textbf{Controlled Scheduler.}
The \texttt{control} and \texttt{replay} modes of \toolname{RunTarget} run on top of the controlled scheduler, a serializing scheduler that we implement following Coyote's approach~\cite{agarwal2021nekara,deligiannis2023industrial}. 
The scheduler takes the instrumented variable accesses and the intercepted synchronization operations as scheduling points. A thread hands control to the scheduler on reaching a scheduling point, and only one thread runs at any moment.
The \texttt{control} mode drives the execution to satisfy the plan $\pi$. Let $s$ denote the length of the satisfied prefix of $S = [e_1 \prec \cdots \prec e_n]$, so the scheduler waits for the next event $e_{s+1}$ and releases it only when its state constraint $\phi_{s+1}$ in $\Phi$ holds, or immediately if $\Phi$ pairs no constraint with $e_{s+1}$. 
At each scheduling point, the scheduler decides following Algorithm~\ref{alg:sched}. The paused threads, each about to perform one event, are partitioned into three classes. A thread in $T_{\mathit{next}}$ is about to perform $e_{s+1}$ with $\phi_{s+1}$ holding, and releasing it advances $s$. A thread in $T_{\mathit{held}}$ is suspended, as its event comes after $e_{s+1}$ in $S$, or is $e_{s+1}$ while $\phi_{s+1}$ does not yet hold, or would change the state of a variable in $\phi_{s+1}$ once $\phi_{s+1}$ holds and before $e_{s+1}$ occurs. The remaining threads irrelevant to the current constraints belong to $T_{\mathit{free}}$ and are scheduled at random. 
When only suspended threads remain or $s$ has not advanced within a timeout, the scheduler forcibly releases a thread or abandons the remaining constraints, and such forced steps do not advance $s$.
A run is therefore biased toward satisfying $\pi$ rather than guaranteed to, and $\kappa_S$ and $\kappa_\Phi$ record how the constraints are actually satisfied in the run, while $\theta$ records the scheduling decision at each scheduling point. 
The \texttt{replay} mode releases threads point by point according to $\theta$ and thus deterministically reproduces the recorded run.

\subsection{Comprehend: Hypothesizing from Runtime Observations and Static Information}
\label{sec:comprehend}
The comprehend phase proposes hypotheses of potential concurrency bugs by combining runtime observations with static program information, and proceeds in four steps.

\emph{Scope measurement.} The agent first calls \toolname{RunTarget} to run the program in \texttt{natural} mode, collects the reached source files, and writes them into \gsname{SCOPE} as the initial scope.

\emph{Runtime observation.} The agent calls \toolname{InstrumentTarget} to instrument the files in \gsname{SCOPE}, then calls \toolname{RunTarget} in \texttt{collect} mode to collect the variable access traces at runtime, and calls \toolname{AnalyzeSharedAccess} to populate \gsname{SHARED\_ACCESS}. From the access profile of each variable, the agent picks the variables of interest, such as those accessed by multiple threads at multiple sites, examines their conflicting access pairs, and judges which pairs serve as clues.

\emph{Static inspection.} Following the runtime clues, the agent reads the relevant source code via \toolname{SearchCode} and calls \toolname{AnalyzeStaticInfo} when needed for call relations and path conditions. Combining the dynamic and static information, the agent reasons about the concurrency semantics: the role of each thread, the state and synchronization operations related to the key shared variables, whether the orders between their accesses are protected, and which reachable code regions remain unexecuted. On this basis, drawing on its knowledge of known concurrency bug patterns, the agent judges which interleavings may break the expected program behavior and with what consequences.

\emph{Hypothesis proposal.} Based on the above analysis and reasoning, the agent proposes a set of bug hypotheses, each of which is written into \gsname{CANDIDATES} as $\langle h, \textsf{pending}, \bot \rangle$.

\textbf{Example 2.}
For the program in Figure~\ref{fig:motivation}, the observed access summary $\sigma_{\mathtt{CodeExists}}$ of the shared variable \texttt{CodeExists} records the events $e_4$, $e_5$, and $e_6$, all with an empty lock set, and $\sigma_{\mathtt{map}}$ records $e_1$, $e_2$, and $e_3$, all with the lock set $\{\mathtt{mutatorsMutex\_}\}$. Through \toolname{AnalyzeSharedAccess}, the agent obtains the conflicting access pair on \texttt{CodeExists}, namely $\langle \mathtt{worker}, \mathsf{W}(\mathtt{CodeExists})@\ell10\rangle$ and $\langle \mathtt{main}, \mathsf{W}(\mathtt{CodeExists})@\ell20\rangle$, where the source file $c$ of the variable access sites is omitted for brevity.
Reading the source code further, the agent understands that \texttt{CodeExists} caches whether \texttt{map} is non-empty, is updated in a different critical section from \texttt{map}, and decides whether \texttt{MemoryRead} performs the bounds check. 
The agent thus proposes the hypothesis $h$ with $V_h = \{\mathtt{CodeExists}, \mathtt{map}\}$, $\mathcal{T}_h = \{\mathsf{W}(\mathtt{CodeExists})@\ell10,\allowbreak \mathsf{W}(\mathtt{CodeExists})@\ell20,\allowbreak \mathsf{R}(\mathtt{CodeExists})@\ell23,$\linebreak $\mathsf{W}(\mathtt{map})@\ell8,\allowbreak \mathsf{W}(\mathtt{map})@\ell17,\allowbreak \mathsf{R}(\mathtt{map})@\ell18\}$ and $\delta$: if a registration occurs between the computation of \texttt{newCount} and the write to \texttt{CodeExists} in \texttt{UnregisterCodeBlock}, the written \texttt{false} overwrites the just-published \texttt{true}, and a subsequent out-of-bounds access skips the bounds check and crashes.

\subsection{Attempt: Guiding Thread Scheduling to Confirm or Refute Bug Hypotheses}
\label{sec:attempt}
The attempt phase confirms the hypothesis of each candidate by guiding thread scheduling to trigger the described bug, or refutes it with code and execution evidence, in the following steps.

\emph{Batch selection.} The agent selects a batch for this round from the \textsf{pending} or \textsf{uncertain} candidates in \gsname{CANDIDATES}. Considering the strength of evidence of each candidate, the number of remaining candidates, and the diversity within the batch, the agent picks the candidates most likely to be triggered, and the batch size is also decided by the agent.

\emph{Subagent dispatch.} The agent dispatches one subagent for each selected candidate, passing the hypothesis $h$ of the candidate, the existing attempts on $h$ in \gsname{ATTEMPTS} if the candidate has been dispatched in previous rounds but remains \textsf{uncertain}, and the budget of the subagent, i.e., at most $p$ plan versions, at most $q$ runs per plan, and the time limit $\Delta_{\mathrm{sub}}$. The output of the subagent is all plans tried in the round with their aggregated execution results, i.e., $[\langle \pi_1, \hat{X}_{\pi_1} \rangle, \langle \pi_2, \hat{X}_{\pi_2} \rangle, \ldots]$. The subagent can use \toolname{SearchCode}, \toolname{AnalyzeStaticInfo}, \toolname{AnalyzeSharedAccess}, \toolname{RunTarget}, and \toolname{AnalyzeExecution}, and can read \gsname{SCOPE} and \gsname{SHARED\_ACCESS}. 
Inside a subagent, the work proceeds in two phases, which alternate until the bug is triggered, refuting evidence is found, or the budget of the subagent is exhausted. 

\begin{itemize}
\item \textbf{Plan.} Based on the hypothesis $h = \langle V_h, \mathcal{T}_h, \delta \rangle$ of the candidate, the subagent works out under what thread schedule the bug described in $\delta$ can be triggered, calling \toolname{SearchCode}, \toolname{AnalyzeStaticInfo}, and \toolname{AnalyzeSharedAccess} when needed. The inference is turned into two kinds of requirements: which events must occur in which order, and whether the relevant variables must be in particular states, corresponding to ordering constraints and state constraints respectively, which form the plan $\pi = \langle S, \Phi \rangle$.

\item \textbf{Examine.} The subagent submits $\pi$ to \toolname{RunTarget} in \texttt{control} mode. The plan is first validated. One that fails is returned for revision and resubmitted without counting toward the $p$ plan versions, and a validated one counts as one version, with at most $p$ versions in total. The controlled scheduler executes each version for at most $q$ runs, and stops early once a run triggers the hypothesized bug. Then the subagent calls \toolname{AnalyzeExecution} to aggregate the results $X_\pi$ of all runs into $\hat{X}_\pi$, and judges the outcome in three cases. If some run is engaged as planned with $\epsilon = \textsf{triggered}$, the bug is triggered. If some run is engaged but $\epsilon \neq \textsf{triggered}$, the interleaving occurred without the expected consequence; the subagent then analyzes whether a constraint is missing or whether no bug exists there. If no run is engaged, the subagent locates the stall from $\hat{X}_\pi$ and adjusts the corresponding constraint. In the latter two cases, the subagent returns to the plan phase and proposes a new version until the bug is triggered, refuting evidence is found, or $p$ or $\Delta_{\mathrm{sub}}$ is reached.
\end{itemize}

\emph{Result collection.} After each subagent returns, the agent updates the status of the candidate in \gsname{CANDIDATES} according to the returned results, attaching the scheduling trace $\theta$ for a \textsf{succeeded} candidate. The agent then writes the attempt $\langle \mathit{round}, A \rangle$ into \gsname{ATTEMPTS}, and adds any new coverage $\gamma$ reported in the execution results to \gsname{SCOPE} and \gsname{SHARED\_ACCESS}.

\textbf{Example 3.}
For $h$ in Example 2, the subagent derives from $\delta$ that the overwrite must follow the worker's publication, and first proposes $\pi_1$ with $S = [e_4 \prec e_6]$ and $\Phi = \emptyset$. In $\hat{X}_{\pi_1}$, every run is engaged yet ends with $\epsilon = \textsf{normal}$. The subagent reasons that the main thread overwrites before the worker enters wasm execution, and the next scheduling point is its next registration, which repairs \texttt{CodeExists} before any read observes it. The subagent then proposes $\pi_2$ with $S = [e_4 \prec e_5 \prec e_6 \prec e_7]$ and $\Phi = \{(e_6, \mathtt{CodeExists} = \mathsf{true})\}$. A run of $\pi_2$ is engaged as planned with $\epsilon = \textsf{triggered}$. The candidate is updated in \gsname{CANDIDATES} as $\langle h, \textsf{succeeded}, \textsf{own} \rangle$ with $\theta$ attached for reproduction, and this round's attempt of $h$, $[\langle \pi_1, \hat{X}_{\pi_1} \rangle, \langle \pi_2, \hat{X}_{\pi_2} \rangle]$, is appended to \gsname{ATTEMPTS}.

\subsection{Assess: Analyzing Results to Steer the Next Round or Conclude}
\label{sec:assess}
After all subagents of a round return, \textsc{Reins} enters the assess phase, analyzes the results of the round as a whole, and accordingly steers the next round or concludes, in the following steps.

\emph{Result propagation.} The agent checks whether the conclusions of this round have implications for the \textsf{pending} or \textsf{uncertain} candidates. Evidence refuting one candidate may also refute those resting on the same mechanism or variable, and executing one candidate may trigger the bug of another. The agent accordingly updates the \textit{status} of such candidates and marks their \textit{settledBy} as \textsf{related}.

\emph{Attribution analysis.} The agent analyzes why the candidates of this round succeed or fail. For the triggered candidates, it summarizes what their hypotheses have in common to infer the likely bug patterns in this program. For the untriggered ones, it distinguishes whether the plans are not engaged or engaged but not crashed to infer why. The patterns of success then bias the batch selection and hypothesis proposal of the next round toward similar candidates and mechanisms, and the causes of failure are passed as hints to the subagents dispatched next.

\emph{Coverage review.} For the files $C_\gamma$ and sites $\mathcal{T}_\gamma$ newly added to \gsname{SCOPE} and \gsname{SHARED\_ACCESS} in the attempt phase, the agent inspects their source and static information, and judges whether the files deserve instrumentation and collection and whether the sites deserve hypothesis proposal.

\emph{Decision.} The agent then decides where the next round starts: from the comprehend phase if the coverage review finds new files or sites worth analyzing, or from the attempt phase if \textsf{pending} or \textsf{uncertain} candidates remain. If all candidates are settled, the whole process ends.

\section{Evaluation}
\label{sec:eval}

To evaluate \textsc{Reins}, we consider the following research questions:
\begin{itemize}[leftmargin=*,labelindent=1.2em,labelwidth=2.2em,
                labelsep=0.6em,align=left,itemsep=1pt,parsep=0pt,topsep=2pt]
  \item[\textbf{RQ1}] How does \textsc{Reins} perform compared with existing controlled scheduling techniques?
  \item[\textbf{RQ2}] Can \textsc{Reins} detect 
        concurrency bugs in real-world projects?
  \item[\textbf{RQ3}] How do the key designs of \textsc{Reins} contribute to its effectiveness?
        
\end{itemize}

\subsection{Experimental Setup}
\label{sec:eval-setup}
\textbf{Implementation.} \textsc{Reins} is implemented on top of the Claude Agent SDK~\cite{claude-agent-sdk}. The six tools described in Section~\ref{sec:gs_tools} are exposed to the agents through an MCP server. The lead agent and the subagents are each driven by a skill: the skill of the lead agent describes the three-phase workflow in Sections~\ref{sec:comprehend}--\ref{sec:assess}, and the skill of the subagent describes how to propose and examine plans for a single candidate in the attempt phase (Section~\ref{sec:attempt}). A subagent is spawned by the lead agent as a separate session that shares the same output directory with the lead agent but has its own context and wall clock, and we set the subagents within a round to be dispatched sequentially.
We use Claude Sonnet~4.6 as the default model. The per-phase parameters are set uniformly as follows: in the comprehend phase, the trace collection in \texttt{collect} mode executes each target 10 times; in the attempt phase, the budget of a subagent is at most $p=5$ plan versions, at most $q=50$ runs per plan, and a time limit $\Delta_{\mathrm{sub}}$ of 30 minutes. The total time budget $\Delta_{\mathrm{total}}$ is set to 5 hours per target throughout the evaluation.

\textbf{Environment.} All experiments are conducted on a machine with a 64-core AMD Ryzen Threadripper PRO 7985WX CPU and 512 GB of memory, running Ubuntu 24.04 LTS.

\subsection{RQ1: Comparison with Existing Techniques}
\label{sec:rq1}
To answer RQ1, we compare \textsc{Reins} with representative controlled scheduling techniques on three publicly available benchmarks of concurrent programs, and evaluate its effectiveness and efficiency.

\textbf{Benchmark.} The three benchmarks are SCTBench, ConVul, and RaceBench.
SCTBench~\cite{thomson2014concurrency,sctbench-repo} comprises buggy pthread programs drawn from test suites of prior concurrency testing and verification work, and we adopt the curated version~\cite{wen2022controlled} of 42 programs. Each program contains a single known bug that manifests as an assertion violation, crash, or hang, and the bugs cover data races, atomicity violations, order violations, and deadlocks. Most programs are under 2 kLOC, and a few link against real-world projects of tens of thousands of lines.
ConVul~\cite{cai2019detecting,convul-repo} consists of 10 bugs extracted from concurrency-related CVEs into single-file programs of under 300 LOC, mostly data races that lead to memory errors such as use-after-free, null-pointer dereference, and double-free.
RaceBench~\cite{liang2023racebench} is an injected concurrency-bug benchmark with built-in assertion oracles, consisting of 15 multi-file programs from PARSEC and SPLASH-2 of about 3 to 26 kLOC. Each program has five variants, and each variant is injected with 20 bugs covering data races, atomicity violations, and order violations, all observable as crashes once triggered.

\textbf{Baseline.} Among the advanced controlled scheduling techniques~\cite{wen2022controlled, kokologiannakis2021genmc, yu2012maple, zhao2025selectively, wolff2024greybox}, we select the state-of-the-art SURW~\cite{zhao2025selectively} and RFF~\cite{wolff2024greybox} as the baselines. In addition, since the scheduler of \textsc{Reins} decides the scheduling points beyond the plans at random, we also include the classic randomized algorithms probabilistic concurrency testing~\cite{burckhardt2010randomized} with depth $d$=3 and $d$=10 (denoted PCT-3 and PCT-10) and random walk (RW)~\cite{thomson2016concurrency}. We use the public implementations of SURW and RFF, and reimplement the PCT-3, PCT-10, and RW algorithms on the same scheduler as \textsc{Reins}.

\begin{figure}[t]
  \centering
  \includegraphics[width=\linewidth]{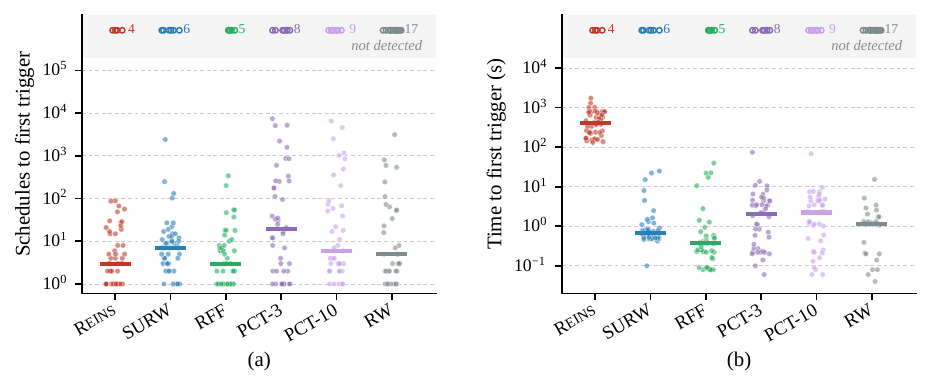}
    \caption{Distribution of the number of schedules (a) and the time (b) to the first trigger on SCTBench and ConVul. Each dot corresponds to one program, and horizontal bars mark the medians over the detected programs; hollow circles in the shaded band, with counts aside, denote the programs not detected within the time budget, and in (b) the lower edge of the band coincides with the budget.}
  \label{fig:rq1-1}
\end{figure}

\textbf{Comparison Results.}
We first evaluate on the 52 programs from SCTBench and ConVul, where each program contains a single bug, and a program is counted as a detection failure if no bug is triggered within $\Delta_{\mathrm{total}}$. Following existing work~\cite{wolff2024greybox,zhao2025selectively}, we measure the number of schedules executed until the bug is first triggered\footnote{For \textsc{Reins}, the number of schedules counts only the executions under the controlled scheduler in the attempt phase.}.
Considering that the tools differ in the granularity of scheduling points and the cost of a single execution, we also report the wall-clock time to the first trigger, as well as the token consumption of \textsc{Reins}.

In terms of detection, \textsc{Reins} detects 48 out of the 52 programs, while SURW, RFF, PCT-3, PCT-10, and RW detect 46, 47, 44, 43, and 35, respectively. The four programs missed by \textsc{Reins} are also missed by all the baselines, and the one additionally missed by RFF is due to its instrumentation not supporting the program.
Figure~\ref{fig:rq1-1}(a) shows that all the detections of \textsc{Reins} are completed within about 100 schedules, whereas the baselines exhibit long tails of $10^3$ to $10^4$ schedules on hard programs. Figure~\ref{fig:rq1-1}(b) shows that \textsc{Reins} is two to three orders of magnitude more expensive than the baselines in time due to the reasoning of the agent, with a median time to the first trigger of about 7 minutes and a maximum of 29 minutes, and its token consumption ranges from 386k/6k to 4.1M/68k input/output tokens per program.
Overall, on these two benchmarks, \textsc{Reins} shows no significant advantage in detection, but the agent guidance brings no degradation, and the far fewer schedules indicate that the constraints derived from the hypotheses do narrow the scheduler down to the interleavings likely to trigger the bug.

We further evaluate on RaceBench, which consists of larger programs. Each program variant is treated as a target and contains multiple injected bugs, and we count the distinct injected bugs triggered within $\Delta_{\mathrm{total}}$.
All tools use the seed input provided by RaceBench, and the triggered bugs are configured not to abort the process on subsequent triggers so that the remaining bugs can be further explored. 
For \textsc{Reins}, since the injected bugs of RaceBench leave recognizable traces in the source code, such as numbered state variables and trigger functions, we mechanically anonymize these traces by renaming the relevant identifiers and removing the injection markers without changing any program instruction, to prevent the agent from locating the injected bugs directly.

\begin{table}[t]
  \centering
  \caption{Results on RaceBench. Each row is a base program with five variants, each injected with 20 bugs (max.~100 per program). kLOC is the average size of the five variants, and each tool column reports the number of distinct injected bugs detected within the time budget. The best result of each row is in bold.}
  \label{tab:rq1-racebench}
  \setlength{\tabcolsep}{4.5pt}
  \begin{tabular}{lrrrrrrr}
    \toprule
    Program & kLOC & \textsc{Reins} & SURW & RFF & PCT-3 & PCT-10 & RW \\
    \midrule
    blackscholes    & 3.0  & 19 & \textbf{32} & 24 & 23 & 25 & 10 \\
    streamcluster   & 3.1  & \textbf{59} & \textbf{59} & 40 & 31 & 43 & 44 \\
    water\_nsquared & 4.0  & 17 & 11 & \textbf{20} &  9 & 12 & 13 \\
    canneal         & 4.1  & 52 & \textbf{55} & 51 & 19 & 27 & 29 \\
    water\_spatial  & 4.6  & \textbf{67} & 41 & 49 & 23 & 33 & 33 \\
    dedup           & 4.8  & 57 & \textbf{58} & 50 & 22 & 24 & 33 \\
    fluidanimate    & 5.1  & \textbf{56} & 55 & 38 & 41 & 46 & 27 \\
    pigz            & 7.2  & 75 & \textbf{76} & 61 & 59 & 62 & 59 \\
    cholesky        & 8.3  & \textbf{65} & 54 & 43 & 32 & 38 & 36 \\
    bodytrack       & 8.4  & \textbf{50} & 38 & 30 & 24 & 26 & 30 \\
    raytrace        & 10.2 & 36 & \textbf{60} & 43 & 24 & 24 & 20 \\
    ferret          & 11.2 & 66 & 67 & \textbf{80} & 41 & 54 & 37 \\
    raytrace2       & 11.4 & 48 & \textbf{49} & 33 & 24 & 24 & 25 \\
    volrend         & 15.2 & \textbf{34} & 31 & 23 & 16 & 17 & 13 \\
    x264            & 26.3 & \textbf{68} & 45 & 60 & 25 & 34 & 30 \\
    \midrule
    Total (max.~1,500) & -- & \textbf{769} & 731 & 645 & 413 & 489 & 439 \\
    \bottomrule
  \end{tabular}
\end{table}

\begin{figure}[t]
  \centering
  \includegraphics[width=0.62\linewidth]{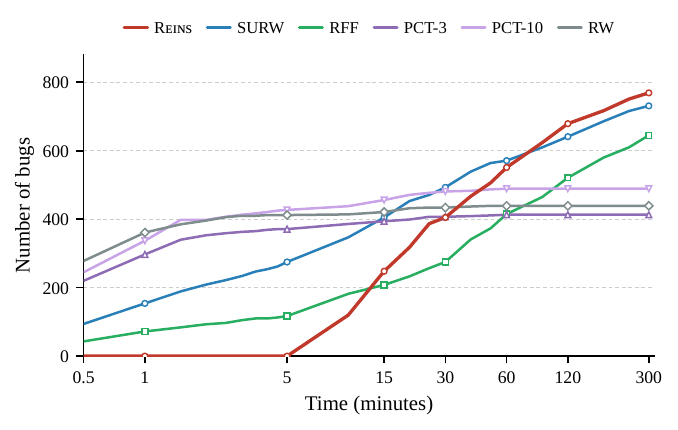}
    \caption{Cumulative number of injected bugs on RaceBench detected over time, summed over the 75 targets.}
  \label{fig:rq1-racebench-time}
\end{figure}

Table~\ref{tab:rq1-racebench} and Figure~\ref{fig:rq1-racebench-time} present the per-program detection and the cumulative detection over time. 
Among the 1,500 injected bugs of the 75 targets, \textsc{Reins} detects 769 bugs, the highest among all the tools and 38 more than the best baseline SURW with 731. 
In terms of the time trend, PCT and RW saturate within about one hour, as random scheduling quickly covers the shallow interleavings. \textsc{Reins} detects no bug in the first five minutes due to the collection and analysis in the comprehend phase, then grows rapidly, surpasses SURW between 60 and 90 minutes, and keeps growing until the budget runs out. 
The end-to-end wall time of \textsc{Reins} ranges from 84 minutes to 5 hours, and the token consumption ranges from 5.7M/68k to 153.1M/912k input/output tokens per target.

Compared with SURW and RFF, \textsc{Reins} performs worse on programs raytrace, blackscholes, and ferret. The untriggered bugs share a common characteristic that the triggering requires the state variables to hold specific values. These values rarely appear at runtime and can only be produced through a large number of interleavings and even the interleavings inside a statement. During the scheduled execution, \textsc{Reins} waits for the values to appear or guides a limited number of interleavings rather than precisely controlling every step of the value accumulation, and its source-level instrumentation misses some interleaving points inside a statement, so the values can hardly appear within the schedule budget of a single plan. As a result, even though the agent correctly understands the triggering conditions, it fails to trigger these bugs. SURW and RFF instrument at the binary level, which allows thread switches inside a statement, and they sample interleavings with a large number of executions, so they can reach these values instead. For \textsc{Reins}, instrumenting the scheduler at a finer granularity or employing a solver to compute the paths to the target values is a feasible improvement.
In addition, all the tools find no more than about 51\% of the 1,500 injected bugs within the 5-hour budget, which comes from two aspects. The first is the test input. RaceBench provides one seed input for each program variant and a separate mutated input that triggers each of the 20 injected bugs. Switching inputs per bug requires dividing the execution budget among multiple inputs with no consistent allocation across tools, and some tools perform additional processing for each input, for example the per-input dynamic preprocessing of RFF over 20 inputs on the variants of cholesky costs time comparable to the whole budget. We therefore uniformly use the seed input, but some injected bugs can only be triggered with the specific data in their mutated inputs, which imposes a common ceiling on the detection of all tools. The second is the time budget. Considering the token cost of \textsc{Reins}, we set $\Delta_{\mathrm{total}}$ to 5 hours, while the curves of \textsc{Reins}, SURW, and RFF are still rising at the end of the budget. In particular, some programs run slowly after instrumentation, and the schedules SURW completes within the budget are far fewer than the 10,000 schedules per target used in its original evaluation.

\begin{answerbox}
\textbf{Answer to RQ1:} Compared with existing controlled scheduling techniques on the three benchmarks, \textsc{Reins} detects more concurrency bugs.
\end{answerbox}

\subsection{RQ2: Bug Finding in Real-world Projects}
\label{sec:rq2}
We further apply \textsc{Reins} to real-world concurrent projects to evaluate its effectiveness.

\textbf{Target Projects.} We select the target projects by tracing the real bugs and base programs in the three benchmarks back to their upstream projects and keeping those still actively maintained, which yields two Mozilla projects, NSPR~\cite{nspr} and SpiderMonkey~\cite{spidermonkey}.
NSPR (Netscape Portable Runtime) is Mozilla's cross-platform system runtime library providing threads, synchronization primitives, and I/O abstractions, with about 136 kLOC. 
SpiderMonkey is Mozilla's JavaScript/WebAssembly runtime featuring JIT compilation and garbage collection, which makes heavy use of multithreading internally and has about 1,214 kLOC.

\textbf{Bug Findings.} \textsc{Reins} detects bugs with given test cases and inputs as program entries. In total, \textsc{Reins} finds six previously unknown concurrency bugs in the two projects, all of which have been reported to the developers via Bugzilla~\cite{bugzilla}. The test cases and detailed bug descriptions are available in our artifact. Table~\ref{tab:rq2-bugs} lists the bugs with aliases assigned in this paper and the scale of code executed in each detection. At the time of writing, the three SpiderMonkey bugs have all been confirmed and fixed by the core developers of the corresponding components. Most of these bugs depend on both specific state preconditions and access orders and are thus hardly triggered in natural runs, and the agent triggers them by guiding the scheduling from its reasoning about the concurrency semantics.

The six bugs cover atomicity violations, a deadlock, and data races. 
Bug 1 is an atomicity violation, where an unlocked read-modify-write on a shared word races with other writers, so that an interrupt request is lost, the waiting thread is never woken up, and the program hangs. 
Bug 2 is a deadlock, where the thread joining the pool and the timer thread acquire the job-queue lock and the timer-queue lock in opposite orders and wait for each other. 
Bug 3 is an atomicity violation, where the enumeration of a wait group caches the base address of the hash table outside the lock while another thread expands the table under the lock and frees the old one, so that the enumeration dereferences freed memory and crashes. 
Bug 4 is a data race, where the main thread overwrites a child slot of a rope string with a parent pointer while flattening it, and the concurrent marking thread reads the slot as a child to mark, crashing on an invalid pointer. 
Bug 5 is a data race, where the main thread inserts into the hash table behind a zone's atom marking bitmap and rehashes it, while the concurrent marking thread looks up the same table and reads an inconsistent or freed table, leading to a crash. 
Bug 6 is another atomicity violation, where two updates that should be performed together are split into two critical sections, and their interleaving leaves a flag stale, so that a wasm out-of-bounds access crashes the process directly instead of being turned into a controlled runtime error. The motivating example in Section~\ref{sec:overview} is a simplified version of this bug.

\begin{table}[t]
  \centering
  \caption{Concurrency bugs found by \textsc{Reins} in real-world projects. AV, DL, and DR denote atomicity violation, deadlock, and data race, respectively. Files and LOC denote the number of source files and lines of code actually executed in the detection run.}
  \label{tab:rq2-bugs}
  \begin{tabular}{llllrrl}
    \toprule
    ID & Name & Project & Type & Files & LOC & Status \\
    \midrule
    1 & \bugname{thread-state}   & NSPR                & AV &  31 &  27k & Reported \\
    2 & \bugname{pool-join}      & NSPR                & DL &  35 &  31k & Reported \\
    3 & \bugname{waitgroup-enum} & NSPR                & AV &  34 &  30k & Reported \\
    4 & \bugname{rope-flatten}   & SpiderMonkey / GC   & DR & 737 & 616k & Fixed \\
    5 & \bugname{atom-bitmap}    & SpiderMonkey / GC   & DR & 829 & 709k & Fixed \\
    6 & \bugname{wasm-codeblock} & SpiderMonkey / Wasm & AV & 909 & 803k & Fixed \\
    \bottomrule
  \end{tabular}
\end{table}
\begin{answerbox}
\textbf{Answer to RQ2:} \textsc{Reins} finds six previously unknown concurrency bugs, three of which have been confirmed by the developers, demonstrating its effectiveness in practice.
\end{answerbox}

\subsection{RQ3: Analysis of Key Designs in \textsc{Reins}}
\label{sec:rq3}
We evaluate the key designs of \textsc{Reins} from three aspects: the role of the runtime observation and static inspection in helping the agent understand the concurrency semantics, the effect of the schedule plans on the controlled scheduler, and the contribution of the iterative mechanisms. We use the five variants of x264 in RaceBench as the targets, with the same setup as in Section~\ref{sec:rq1}.

\textbf{Runtime Observation and Static Inspection.} To evaluate the two designs in the comprehend phase, we construct two ablation variants.
w/o Runtime removes the runtime observation, so that the agent proposes candidates by reading the source code and invoking static analysis, without the observed access summaries as clues.
w/o Static removes the static inspection, so that the agent proposes candidates directly from the collected shared accesses and conflicting access pairs without reading the source code or invoking static analysis.
In addition, we evaluate General Agent as a baseline, a vanilla agent on the Claude Agent SDK that has only the format of schedule plans and the controlled scheduler as the execution infrastructure, and detects bugs on its own by reading the source code and analyzing the program.
Table~\ref{tab:rq3-ablation} reports the total numbers of detected bugs, proposed candidates, and dispatched subagents over the five variants, as well as the dispatch effectiveness and the average time and token cost. In the attempt phase, a subagent is dispatched for each selected candidate to examine it by execution, which we call a \textit{dispatch}, while some candidates are settled directly in the assess phase. A dispatch is effective if the subagent triggers at least one bug with its plans, and Disp.\ Eff.\ is the fraction of effective dispatches.

The full \textsc{Reins} detects the most bugs. The two ablation variants propose and dispatch similar numbers of candidates to \textsc{Reins}, so their gaps mainly lie in the quality of the hypotheses.
w/o Runtime drops the most. Without runtime clues, the agent has to search the code blindly for shared variables, which raises the token consumption, and its candidates largely deviate from the actually executed paths, reducing the effectiveness from 17.6\% to 6.6\%. Their plans often stall at events that never occur, so the subagents end early once they confirm from the code that these events are unreachable under the given input, and the refuting evidence settles the related candidates through result propagation. As a result, all candidates are settled after 182 minutes on average, before the budget runs out. Its detection is even lower than that of General Agent, which has no candidate mechanism, is free to change its exploration direction, and uses up the whole budget.
For w/o Static, the candidates come from the observed conflicting access pairs, many of which are benign races, so its effectiveness decreases to 11.8\%. The full \textsc{Reins} rules out such pairs by static inspection in the comprehend phase, whereas w/o Static defers this judgment to each subagent, which examines its pair by reading the code, invoking static analysis, or executing it. This raises the cost of each dispatch, so the token consumption nearly doubles and the runs take nearly the whole budget.
General Agent consumes the fewest output tokens but detects only 68\% of the bugs of \textsc{Reins}, since its unstructured exploration is slow and shallow.
Note that even the full \textsc{Reins} has an effectiveness of only 17.6\%, but the dispatches that trigger no bug are not wasted, as a refuted candidate rules out a class of possible bugs, and the execution results are used for result propagation and attribution analysis to guide later rounds.

\begin{table}[t]
  \centering
 \caption{Results of \textsc{Reins} and the compared configurations on the five variants of x264. \#Bugs (max.\ 100), \#Cand., and \#Disp.\ are the total numbers of detected bugs, proposed candidates, and dispatched subagents; Disp.\ Eff.\ is the fraction of dispatched subagents that trigger at least one bug; Avg.\ Time and Avg.\ Token are the average time and input/output token consumption per target.}
  \label{tab:rq3-ablation}
  \begin{tabular}{lrrrrrr}
    \toprule
    & \#Bugs & \#Cand. & \#Disp. & Disp.\ Eff. & Avg.\ Time & Avg.\ Token \\
    \midrule
    \textsc{Reins} & 68 & 453 & 159 & 17.6\% & 240 min &  62.5M/526k \\
    w/o Runtime    & 41 & 397 & 137 &  6.6\% & 182 min &  79.4M/603k \\
    w/o Static     & 56 & 389 & 153 & 11.8\% & 297 min & 118.4M/981k \\
    General Agent  & 46 &  -- &  -- &     -- & 300 min &  72.2M/156k \\
    RW             & 30 &  -- &  -- &     -- & 300 min &          -- \\
    \bottomrule
  \end{tabular}
\end{table}

\textbf{Schedule Plans.} The schedule plans bridge the agent's reasoning and the scheduler's execution, as the scheduler follows them at the relevant scheduling points and keeps the rest random. We therefore compare the agent-guided configurations against RW, which runs on the same scheduler without guidance. As shown in Table~\ref{tab:rq3-ablation}, all the agent-guided configurations detect more bugs than RW, and the full \textsc{Reins} more than doubles the detection of RW while covering all of its bugs. The plans thus effectively direct the scheduler toward bug-triggering interleavings.

\begin{figure}[t]
  \centering
  \includegraphics[width=\linewidth]{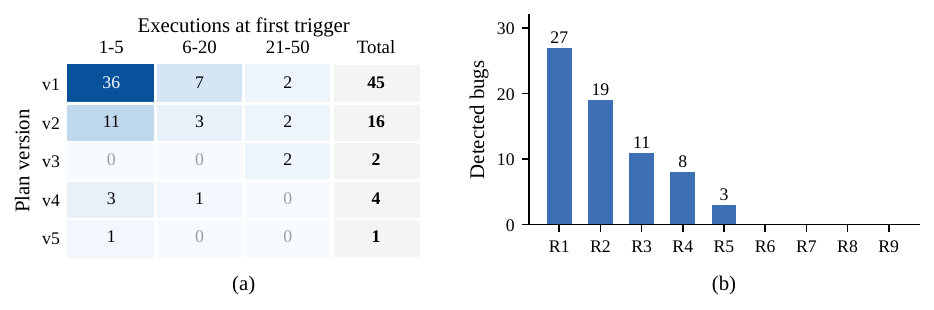}
  \caption{Iteration in \textsc{Reins} on the five variants of x264. (a) Distribution of the 68 triggered bugs over the plan version and the number of executions at the first trigger. (b) Number of bugs detected in each round, summed over the five targets by round number, where x264\_1 to x264\_5 run 8, 7, 3, 9, and 3 rounds, respectively.}
  \label{fig:rq3-iteration}
\end{figure}

\textbf{Iterative Mechanisms.} The iterative mechanisms feed the execution results back into subsequent attempts at two levels. Within the attempt phase, a subagent revises its plan according to the execution results for at most $p$ versions. Across rounds, the assess phase consolidates the results of a round to steer the next one.
As shown in Figure~\ref{fig:rq3-iteration}(a), among the 68 bugs, 45 are triggered by the first plan version and 23 by a version revised according to the execution results. Meanwhile, 91\% of the bugs are triggered within the first 20 executions of a version, and the fourth and fifth versions contribute only 5 in total. Thus, once the plan captures the required interleaving, little scheduling exploration is needed, and the budget of $p=5$ versions and $q=50$ executions is sufficient.
As shown in Figure~\ref{fig:rq3-iteration}(b), 27 bugs are triggered in the first round and the remaining 41 in later rounds, owing to the maintenance of the candidate pool by the assess phase. The five targets produce 453 candidates in total, far more than a single round can verify, and only 159 of them are dispatched (Table~\ref{tab:rq3-ablation}). The assess phase selects the candidates most worth verifying in the next round according to the attribution of the current results, and settles some directly by result propagation. It also expands the pool in two ways, either returning to the comprehend phase to instrument and collect the newly reached files, or directly proposing new hypotheses on the previously accumulated sites. Both occur on x264, each sharply expanding the pool, and these candidates enter verification only through the iteration.

\begin{answerbox}
\textbf{Answer to RQ3:} Runtime observation and static inspection jointly help the agent build the semantic understanding, the schedule plans direct the scheduling and thus improve detection, and the iteration within an attempt and across rounds yields more detections.
\end{answerbox}

\section{Discussion}
\label{sec:discussion}
\textbf{Cost and Efficiency.}
The time cost of \textsc{Reins} mainly comes from the reasoning of the agent, and about 87\% of the wall time is spent in the subagents. To keep the same execution parallelism on the controlled scheduler as the baselines and to stay within the usage limits of the model service, the subagents within a round are executed sequentially in our experiments; however,  parallel execution is possible and it would significantly reduce the wall time.
As for the token cost, about 90\% of the input tokens reported in the evaluation are prompt cache reads, which are priced far below regular input tokens, and the subagents, whose tasks are simpler than those of the lead agent, can be driven by cheaper models.
In addition, \textsc{Reins} can be combined with traditional techniques, letting cheap random scheduling pick up the shallow bugs first and leaving the hard ones to the agent, as supported by the finding in Section~\ref{sec:rq3} that all 30 bugs detected by RW are included in the 68 bugs of \textsc{Reins}. In future work, we will explore these ideas to improve the efficiency of \textsc{Reins} while retaining its detection capability.

\textbf{Threats to Validity.} The randomness of LLM outputs is the main threat, as the hypotheses and plans generated by the agent may differ across runs. Due to the high cost of each run, each tool and configuration is run once, and the agent uses a single model. Our evaluation on 127 targets with more than 1,500 bugs mitigates this risk, as the conclusions are drawn from the overall results rather than from individual targets. 
Data leakage is another concern. The injected bugs in RaceBench leave traces in the source code, which we anonymize to prevent the agent from locating the bugs directly. The benchmark code may appear in the training corpus of the model, but memorization at most helps locate suspicious code and cannot replace constructing a replayable schedule that triggers the bug. General Agent with the same model detects far fewer bugs than \textsc{Reins}, indicating that the detection depends mainly on the workflow design rather than pure memorization. In addition, the previously unknown bugs found in real-world projects are free from data leakage. Regarding the comparison with the baselines, the original papers of some baselines use the number of schedules as the budget, whereas we evaluate all tools under the same time budget, so our results are not directly comparable with those reported in the original papers, while the comparison under the same budget remains valid.

\textbf{Limitations.}
First, like other CCT techniques, \textsc{Reins} detects bugs only in the code reached by the given test entries and inputs. Second, the current implementation supports C/C++ programs with shared-memory multithreading, and extending it to other languages or concurrency models requires adapting the instrumentation and scheduling layers. Third, \textsc{Reins} detects only the bugs that manifest as crashes, assertion failures, or hangs, and those with silent symptoms such as data corruption require additional oracles.

\section{Related Work}
\textbf{Concurrency Testing.}
Many program analysis~\cite{pratikakis2011locksmith, blackshear2018racerd, cai2021canary, flanagan2009fasttrack, serebryany2009threadsanitizer, zhu2023tolerate}, verification~\cite{kokologiannakis2021genmc, godefroid1996partial, clarke2004tool}, and testing~\cite{edelstein2002multithreaded, xu2020krace, thomson2016concurrency} techniques have been proposed for concurrency bug detection and reproduction.
In this work, we mainly focus on controlled scheduling techniques.
Controlled concurrency testing explores the interleaving space systematically or randomly.
Systematic concurrency testing stems from stateless model checking~\cite{godefroid1997verisoft, flanagan2005dynamic, kokologiannakis2021genmc} and enumerates the interleaving space under certain bounds, such as the preemption bound~\cite{musuvathi2007iterative, musuvathi2008finding} and the delay bound~\cite{emmi2011delay}, while \textsc{Period}~\cite{wen2022controlled} divides the execution into periods and explores their interleavings guided by the execution history.
Randomized concurrency testing, represented by the random priority scheduling of \textsc{PCT}~\cite{burckhardt2010randomized}, samples the interleaving space randomly with a probabilistic guarantee of triggering bugs. \textsc{QL}~\cite{mukherjee2020learning} learns scheduling decisions from past executions with Q-learning to bias the scheduler toward bug-exposing choices. 
\textsc{RFF}~\cite{wolff2024greybox} applies greybox fuzzing, partitioning the search space by the reads-from relation of executions and pruning the covered partitions to steer exploration toward the under-explored ones. 
\textsc{SURW}~\cite{zhao2025selectively} targets the interleaving uniformity missed by prior randomized algorithms and samples interleavings uniformly over a selected subset of events.
These techniques decide the schedules from bounds, probabilities, or execution signals at the level of events and executions, whereas \textsc{Reins} guides thread scheduling with the agent's reasoning about the program semantics.
In addition, there exist some studies that combine predictive analysis with active thread scheduling to detect bugs of specific patterns, known as active testing~\cite{joshi2009calfuzzer}.
These studies predefine an interleaving pattern for each type of bug, predict its potential instances composed of a few program events with their order or dependencies, and actively schedule threads to trigger them.
For example, \textsc{RaceFuzzer}~\cite{sen2008race} detects data races with potentially racing statement pairs, \textsc{CTrigger}~\cite{park2009ctrigger} detects atomicity violations with unserializable interleavings of two local accesses and one remote access, and \textsc{DeadlockFuzzer}~\cite{joshi2009randomized} detects deadlocks with potential cycles of lock dependencies.
\textsc{Maple}~\cite{yu2012maple} further generalizes these patterns into idioms of one or two inter-thread dependencies and exposes their untested instances by active scheduling.
By comparison, \textsc{Reins} is not limited to any specific or predefined interleaving pattern. Its hypotheses describe the thread interactions that may go wrong based on the program semantics, its schedule plans constrain a variable number of events and can carry conditions on program states, and the execution results are fed back to subsequent attempts to revise the plans or hypotheses.

\textbf{LLM Agents for Software Testing.} 
LLMs and agents have recently been applied to a wide range of software testing tasks~\cite{wang2024software}, such as unit test case generation~\cite{wang2024hits}, test oracle generation~\cite{dinella2022toga}, test input generation~\cite{xia2024fuzz4all, xu2025ckgfuzzer, zhu2026locus}, and bug analysis~\cite{kang2023large, lee2026agentic, pu2026patch, simsek2026pocgen}.
Recent work couples the agent with program analysis and dynamic validation in an iterative loop, such as \textsc{Locus}~\cite{zhu2026locus} synthesizing progress predicates for directed fuzzing, \textsc{ConcoLLMic}~\cite{luo2026agentic} performing symbolization and constraint solving in natural language for concolic testing, and \textsc{K-Repro}~\cite{pu2026patch} and \textsc{PoCGen}~\cite{simsek2026pocgen} generating and validating PoCs for vulnerability reproduction. 
On the concurrency side, \textsc{Slyp}~\cite{lee2026agentic} uses a ReAct-style agent with binary-exploration and debugging tools to locate race-induced memory corruption in COTS binaries and synthesizes concurrent-invocation PoCs that are validated by native execution, whereas \textsc{Reins} triggers bugs by taking over and guiding thread scheduling. 
\textsc{ConCovUp}~\cite{cai2026concovup} generates multithreaded test drivers for concurrency testing with multiple collaborating agents, identifying conflicting shared-memory access pairs through static analysis and deducing the inputs to reach them to improve the coverage of shared-memory access pairs. It focuses on test driver generation, while \textsc{Reins} focuses on scheduling control over the covered code, making the two complementary. 


\section{Conclusion}
This paper presents agent-guided thread scheduling and implements it in \textsc{Reins}, which detects more bugs than existing controlled scheduling techniques on three public benchmarks and finds six previously unknown concurrency bugs in two real-world projects. 
These results demonstrate that the agent's reasoning about the concurrency semantics can guide thread scheduling to trigger concurrency bugs more effectively.


\bibliographystyle{plain}
\bibliography{references}

\end{document}